\documentclass[]{article}
\usepackage{epsf,a4}
\begin{document}
\begin{center}
\LARGE
Parallel Sparse Matrix Multiplication for \\
Linear Scaling Electronic Structure Calculations
\bigskip \\
\large
D. R. Bowler$^\star$, T. Miyazaki$^\dagger$$^\star$ and 
M. J. Gillan$^\star$ \smallskip \\
$^\star$Department of Physics and Astronomy, University College London \\
Gower Street, London, WC1E 6BT, U.K. \smallskip \\
$^\dagger$National Research Institute for Metals \\
1-2-1 Sengen, Tsukuba, Ibaraki 305-0047, Japan
\normalsize
\end{center}

\begin{abstract}
Linear-scaling electronic-structure techniques, also called $O(N)$
techniques, rely heavily on the multiplication of sparse matrices,
where the sparsity arises from spatial cut-offs.  In order to treat
very large systems, the calculations must be run on parallel
computers. We analyse the problem of parallelising the multiplication
of sparse matrices with the sparsity pattern required by
linear-scaling techniques. We show that the management of inter-node
communications and the effective use of on-node cache are helped by
organising the atoms into compact groups. We also discuss how to
identify a key part of the code called the `multiplication kernel',
which is repeatedly invoked to build up the matrix product, and
explicit code is presented for this kernel. Numerical tests of the
resulting multiplication code are reported for cases of practical
interest, and it is shown that their scaling properties for systems
containing up to 20,000 atoms on machines having up to 512 processors
are excellent. The tests also show that the cpu efficiency of our code
is satisfactory.
\end{abstract}

\section{Introduction}
\label{sec:intro}

There is rapidly growing interest in linear-scaling methods for
electronic structure calculations, in which the memory and number of
cpu cycles are proportional to the number of atoms\cite{Goedecker99}.
Several practical codes exist for performing
tight-binding~\cite{Pettifor89,Yang91,%
Galli92,Li93,Daw93,Ordejon93,Aoki93,Mauri93,Goedecker94,%
Kress95,Kohn96,Horsfield96,Bowler97},
first-principles~\cite{Stechel94,Hernandez95,Haynes97} or
Hartree-Fock~\cite{Schweg96,Burant96,Ochsen98,Challa00} calculations
using linear-scaling techniques -- often called order-$N$ or $O(N)$
techniques. For very large systems containing thousands or tens of
thousands of atoms, these codes need to run efficiently on parallel
computers. Most of the existing linear-scaling methods rely heavily on
the multiplication of sparse matrices, and this means that
sparse-matrix multiplication on parallel computers is a key issue for
the future of linear-scaling electronic structure work. To achieve the
best efficiency, parallel codes to perform this multiplication must be
designed to treat the special patterns of sparsity that arise in
electronic structure, which we shall refer to as `local sparsity'.
The aims of this paper are to analyse the design of parallel code for
multiplying matrices having local sparsity, to present the algorithms
we have developed for the {\sc Conquest} electronic structure
code~\cite{Goringe97,Bowler00}, and to report the results of practical
tests on the efficiency and linear-scaling properties of the code.

In the tight-binding approach to electronic structure, we are
concerned with the Hamiltonian and overlap matrices $H_{i \alpha , j
\beta}$ and $S_{i \alpha , j \beta}$. (Here, $i$, $j$ label atoms,
while $\alpha$, $\beta$ label the basis functions on each atom.)
Linear-scaling methods generally require also the density matrix $K_{i
\alpha , j \beta}$ and sometimes also an `auxiliary' density matrix
$L_{i \alpha , j \beta}$ -- we follow the notation of
Ref.~\cite{Hernandez96}. Linear-scaling methods based on
density-functional theory (DFT) or Hartree-Fock theory work with
similar quantities, and the DFT-pseudopotential approach also needs
scalar products $P_{i \alpha , j \lambda}$ between localised orbitals
on atom $i$ and angular-momentum projectors for non-local
pseudopotentials on atom $j$.  The elements of all these matrices tend
to zero as the distance $R_{i j}$ between atoms $i$ and $j$ tends to
infinity, and this is what makes linear-scaling eletronic structure
theory possible.  In the approach used in the {\sc Conquest}
code~\cite{Goringe97,Bowler00}, and other related approaches (see
e.g. Refs~\cite{Ordejon93,Mauri93, Mauri94,Ordejon95}), the
approximation is made that all the matrix elements vanish exactly when
$R_{i j}$ exceeds a specified cut-off radius, which generally differs
for different matrices. Once this approximation has been made,
linear-scaling behaviour for both memory and cpu cycles is guaranteed
-- provided, of course, appropriate algorithms are used. All the
matrices are sparse, since their only non-vanishing elements are those
for which $R_{i j}$ is less than the appropriate cut-off. In
everything that follows, the particular pattern of sparsity that
arises from the spatial cut-offs is crucial, and it is this pattern
that we call `local sparsity'.

The techniques widely used to ensure the idempotency of the density
matrix $K$~\cite{Li93,Hernandez96,Bowler99} rely on the calculation of
matrix products such as $(LSL)_{i \alpha , j \beta}$, $(LSLSL)_{i
\alpha , j \beta}$ and other similar products involving $H$ and
$P$. Sometimes, these products are needed only for interatomic
distances $R_{i j}$ that are smaller than the full distances for which
they have non-vanishing elements, so that we do not need to calculate
all their elements.  The general problem is therefore to perform the
product:
\begin{equation}
C_{i \mu , j \nu} = \sum_{k \xi} A_{i \mu , k \xi}
B_{k \xi , j \nu} \; ,
\label{eqn:mult}
\end{equation}
where the elements of $A$ and $B$ are non-vanishing only
for $R_{i k} < R_A$ and $R_{k j} < R_B$ respectively, and elements
of $C$ are required only for $R_{i j} < R_C$, where $R_A$,
$R_B$, $R_C$ are specified cut-off distances. The indices
$\mu$, $\nu$, $\xi$ may correspond to localised orbitals or 
angular momentum projectors
on each atom, but in any case they run only over a small set
of values. In general, the number of values might depend on
the atom, but we ignore this possible complication here and assume that
$\mu$ runs from 1 to $n_1$, $\nu$ runs from 1 to $n_2$, and $\xi$
runs from 1 to $n_3$. Unless the distinction is important,
we shall refer to $n_1$, $n_2$ or $n_3$ simply as $n$.

The {\sc Conquest} DFT code was written from
the start as a parallel code, and the methods used to parallelise
sparse-matrix multiplications and all the other operations
were reported earlier~\cite{Goringe97}, together with
tests of the scaling behaviour. Although the scaling
was excellent, the major problem
with the matrix multiplication methods that we used earlier
was that they required elaborate indexing, which consumed
excessive amounts of memory. This is why we have carried out
the much more thorough analysis of locally sparse matrix
multiplication reported here.

There has also been other work on parallelising the multiplication
of locally sparse matrices for linear-scaling electronic structure
work, using both tight-binding~\cite{Itoh95,Canning96,Wang96} 
and Hartree-Fock
techniques~\cite{Challa00}. Itoh {\em et al.}~\cite{Itoh95} and
Canning {\em et al.}~\cite{Canning96} describe the parallelisation
of the two closely related linear-scaling methods reported
in Refs.~\cite{Ordejon93,Ordejon95} and~\cite{Mauri93,Mauri94}, 
and find satisfactory
speed-ups of up to 7.3 for an eight-fold increase of
processor number. Challacombe~\cite{Challa00} presented a very general
parallelisation scheme for linear-scaling Hartree-Fock
calculations using Gaussian basis sets, but found speed-ups
of only 75~\% for an eight-fold increase of processor number.
Wang {\it et al.} made passing reference to the parallelisation
of the tight-binding density-matrix method~\cite{Wang96}, but no details
appear to have been published. The popular tight-binding
linear-scaling scheme known as the Fermi-operator 
method~\cite{Goedecker94}
has also been parallelised, but this relies on matrix-vector
rather than matrix-matrix operations, and is less relevant here.

In considering the efficiency of a multiplication code, we note that
there is a certain irreducible minimum number of multiply-and-adds
that must be done to accomplish the calculation of
eqn~(\ref{eqn:mult}).  For a parallel machine whose processors have a
given peak speed, this sets a well defined lower bound to the
execution time.  The ratio between this lower bound and the 
practical execution time gives a measure of the cpu efficiency of the
code. The superscalar processors of main interest to us can sometimes
achieve 50~\% of peak speed in the multiplication of large matrices,
but it would be unrealistic to expect this kind of efficiency for
sparse matrices. On the other hand, it would be cause for concern if
the cpu efficiency fell much below 10~\%. We shall show that
efficiencies of around 10~\% are indeed achievable.  We shall show
that the methods we propose achieve scaling very close to linear, both
as the numbers of atoms and processors are increased for a fixed
number of atoms per processor, and as the number of processors is
increased for a fixed number of atoms.

We shall see that achievement of high efficiency and good scalability 
relies heavily on two things: first, good design of storage patterns
of matrix elements; second, the use of different schemes for
labelling atoms in different parts of the calculation, together with
efficient transcription between schemes. For both storage and labelling,
we shall show the advantages of grouping atoms into small spatially
compact groups (here called `partitions'), and assignment of groups
of these partitions (here called partition `bundles') to individual
processors.

The next section of the paper presents our analysis of the problem,
and the set of algorithms that the analysis leads us to. We also
present the piece of code that lies at the heart of our multiplication
program, which we refer to as the `multiplication kernel'.  Our
practical code is written in Fortran~90, with communications handled
by MPI1. In Sec.~3, we outline the main principles used in
programming our algorithms. Sec.~4 reports the range of tests on
a Cray~T3E, an SGI Origin2000 and a beowulf cluster
that we have performed to probe the efficiency and scaling
behaviour of the multiplication code.  The final Section discusses our
results and draws conclusions. In particular, we shall argue that
other possible approaches to the problem of multiplying locally sparse
matrices on parallel machines are unlikely to achieve significantly
better efficiency or linear scaling behaviour.

\section{Development of techniques}
\label{sec:techniques}

In order to achieve high efficiency, we need careful design
of a small set of operations that are repeatedly performed
to build up the product matrix: this set of
operations is the `multiplication kernel'. Some of the
data that serve as input to the kernel need to be communicated
from remote nodes, and linear scaling will only be achieved
if the communications strategy is properly designed. Both the
multiplication kernel and the communications need indexing,
and it is important that both the memory 
and the cpu time needed to manipulate index arrays are kept to a
minimum. Before discussing the multiplication kernel, the
communications strategy and the indexing, we outline first
the starting assumptions and the spatial organisation of
atoms into partitions and bundles.

For ease of presentation, we confine ourselves in much of the
following to the case $R_C \ge R_A + R_B$, which we
refer to as the `maximal' case; it is the case where every
element of the product matrix $C = A \cdot B$ must be calculated. 
At the end of this Section, we shall
show how the case of a general relation between $R_A$, $R_B$ and
$R_C$ can be handled by exactly the same principles that will
be described for the maximal case. We also avoid initially any
discussion of periodic boundary conditions. The {\sc Conquest}
code is able to handle periodic boundary conditions, though 
this may not always be needed, so that the matrix multiplication
code must be capable of including periodicity.
But to simplify the discussion,
we defer discussion of periodicity till Sec.~\ref{sec:pbc}.

\subsection{Assumptions}
\label{sec:assumptions}

Following Ref.~\cite{Goringe97}, we assume that each node is
responsible for a group of atoms: we call this the node's `primary
set' of atoms.  We also assume that matrix elements are stored by
rows, so that for any matrix $X$, the elements $X_{i \mu , j \nu}$ for
all $\mu$, $j$, $\nu$ are held by the node whose primary set contains
$i$.  Finally, we assume that the multiplication operations needed to
calculate the elements $C_{i \mu , j \nu}$ of the product matrix are
performed by the node whose primary set contains $i$.  This means that
only matrix elements $B_{k \xi , j \nu}$ need to be communicated
between nodes. We discuss in Sec.~\ref{sec:disc} whether anything
could be gained by changing these assumptions.

With these assumptions, we already encounter a key question, concerning
the interleaving of communications and calculations. The matrix
multiplications performed by each node can be broken into
contributions, with each contribution associated with
a particular set of atoms $k$. For each set, we fetch the
$B_{k \xi , j \nu}$ data for $k$ in the set, and then perform
the multiply-and-adds of eqn~(\ref{eqn:mult}) for those $k$, accumulating the
results onto the array of the product matrix; then we move to the next
set of $k$ atoms and repeat the communications and calculations.
Two extremes can be envisaged. At one extreme, we start by
bringing {\em all} the $B_{k \xi , j \nu}$ for all the $k$
entering eqn~(\ref{eqn:mult}) from the remote nodes onto the local node;
after this has been done, we then perform {\em all} the multiplications.
This is the coarsest possible interleaving. At the other extreme, we
fetch the $B_{k \xi , j \nu}$ for just a single atom
$k$ at a time, and perform the multiplications for that $k$, before
moving on to the next $k$. This is the finest possible interleaving.
Neither extreme will be satisfactory. The coarse extreme requires
an unnecessarily large amount of local memory to store all the
$B_{k \xi , j \nu}$ elements; the finest extreme requires
the transmission of unnecessarily small packets of
$B_{k \xi , j \nu}$ data, so that latency will slow the
communications. Somewhere between the extremes, there is a good
compromise between memory and latency.

In fact, there is no point in communicating {\em all}
$B_{k \xi , j \nu}$ data before multiplying: the coarsest interleaving
that we should ever consider is where we communicate all the
$B_{k \xi , j \nu}$ needed from a particular remote node before
performing the multiplications; then we move to the next node.
This interleaving already yields the lowest possible latency. It follows
that the outermost loop in matrix multiplication must be a loop
over remote nodes. However, there is no reason why this coarsest
interleaving should give an acceptable compromise, and we shall
generally want to reduce the memory requirement further by
breaking the atoms $k$ into smaller sets. Within the outer loop
over nodes, we shall therefore want a loop over {\em sets} of
atoms $k$ in each particular node.

\begin{figure}
\begin{center}
\leavevmode
\epsfxsize=100mm    
\epsfbox{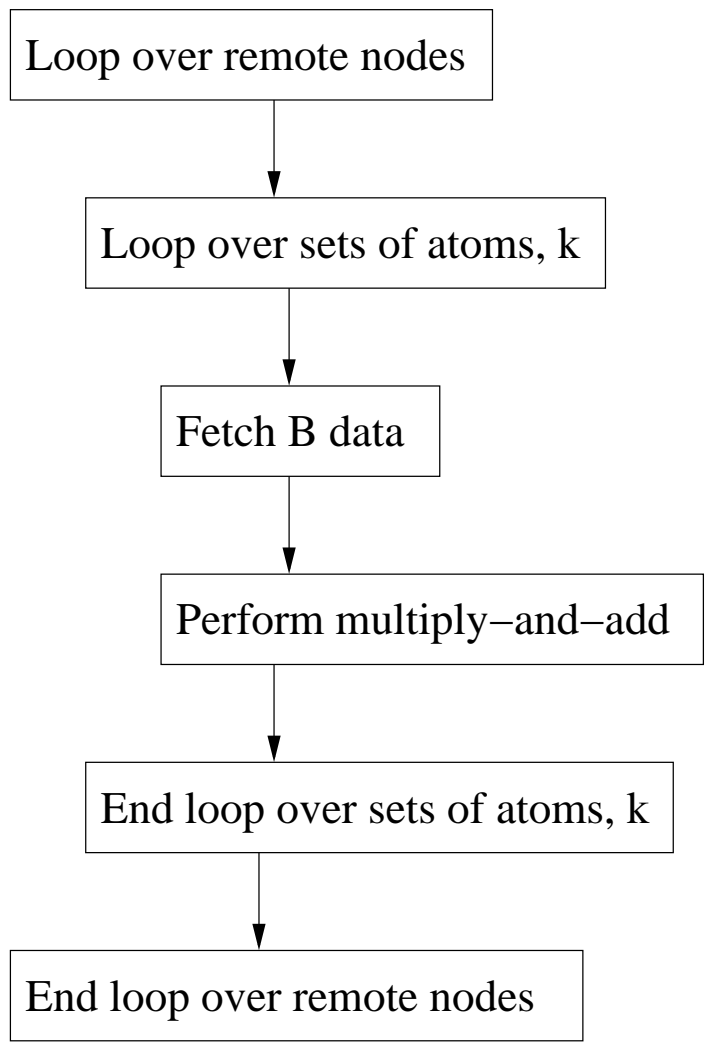}
\end{center}
\caption{Interleaving of communication and calculation in 
parallel multiplication of sparse matrices.}
\label{fig:interleave}
\end{figure}

Our basic assumptions about the distribution of matrix elements
over nodes therefore require the general interleaving of communications
and calculations shown in Fig.~\ref{fig:interleave}.

\subsection{Partitions, bundles, haloes and the outline
multiplication scheme}
\label{sec:partitions}

The primary sets of atoms should be chosen to be spatially
compact: the atoms should all be near each other. The reason for
this is that compactness of the primary set helps to
reduce the number of atoms $k$ entering equation~(\ref{eqn:mult}) for any
given node, so that fewer $B_{k \xi , j \nu}$ elements have
to be communicated. Put another way, primary set compactness means that
each $B_{k \xi , j \nu}$ communicated is used more times.
Since the time spent in communications limits the parallel
scalability, compactness help scalability.

However, it would limit flexibility if we took the primary sets
to be the smallest organisational unit of atoms. The size
of the primary sets depends on the number of processors used, and this
may depend on unpredictable circumstances. So we need a smaller and
more stable organisational unit. We call this a `partition'. A partition
is a small, spatially compact set of atoms that does not depend on the
number of processors. Each primary set consists of a spatially
compact bundle of partitions. Clearly, in assembling partitions
to make primary sets, a key consideration will be load balancing,
so we shall need each primary set to contain roughly the same
number of atoms. Load balancing will be discussed in more detail
in Sec.~\ref{sec:load}.

Partitions could in principle be formed in many ways. At present,
the simulation cell used in the {\sc Conquest} code is required
to be orthorhombic (this requirement will, of course, be
removed in the future). We divide each of the three edges
of this cell using a uniform grid, and the orthorhombic subcells thus
formed are used as partitions. For some arrangements of atoms -- for
example, slabs of crystal surrounded by vacuum -- there may be
partitions that contain no atoms. This creates no problem,
though for good load balancing we shall require that no primary sets
should consist only of empty partitions.

\begin{figure}
\begin{center}
\leavevmode
\epsfxsize=100mm    
\epsfbox{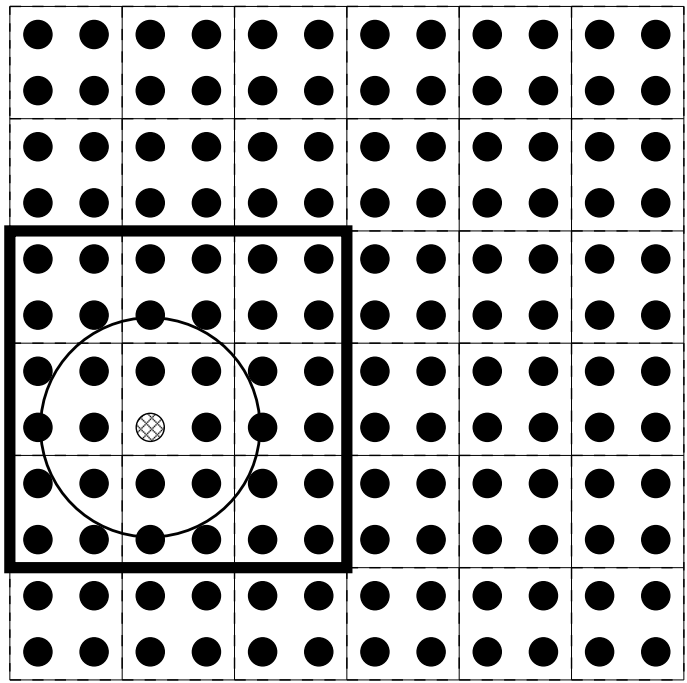}
\end{center}
\caption{Schematic illustration of partitions and haloes.  Atoms are 
indicated by small, black circles, and partitions by the dashed lines.
A particular atom, $i$, is patterned.  Its A-neighbours are those atoms
that are within or touched by the circle drawn centred on $i$.  
If the primary set is chosen to be the partition containing $i$, then the
A-halo partitions are those enclosed by the heavy line.}
\label{fig:halo}
\end{figure}

To discuss the outline multiplication scheme in terms of partitions,
we now introduce some terminology, which is illustrated schematically
in Fig.~\ref{fig:halo}. For each atom $i$ in the
primary set, there are atoms $k$ in the system
whose distance from $i$ is less
than $R_A$; we call these the $A$-neighbours of $i$. The atoms
$k$ which are $A$-neighbours of at least one atom in the {\em primary
set} form a set which we call the $A$-halo. We refer to atoms in this
set as the $A$-halo atoms. The set of {\em partitions} containing at least one
$A$-halo {\em atom} are called the $A$-halo {\em partitions}. Finally,
the set of nodes responsible for at least one $A$-halo partition are
called the $A$-halo {\em nodes}.

The multiplication scheme can now be expressed thus: We loop over
$A$-halo nodes. For each such node, we loop over groups of atoms $k$,
communicate the $B_{k \xi , j \nu}$ elements for atoms in the group,
perform the corresponding multiplications, and accumulate onto
the $C$-matrix array. The groups of atoms $k$ will now be identified as
$A$-halo partitions, so that the inner 
loop in Fig.~\ref{fig:interleave} is a loop over the
$A$-halo partitions on the given $A$-halo node.
Partitions therefore serve two purposes: first, they aid in the
flexible construction of primary sets; second, they provide
the sub-units needed in the communication of $B$-matrix data.

A number of issues are raised by the partition scheme. We must
consider the size of $B$-data packets to be communicated, and how to avoid
fetching unwanted $B$-data. This question is intimately
related to the choice and structure of the multiplication
kernel. It is also related to the question
of the indexing needed so that the multiplication kernel
knows which atom
pairs are associated with each piece of $B$-data. Finally, we must
consider the indexing needed by the multiplication kernel in order to
accumulate $C$. Since the nature of the multiplication kernel lies at the
heart of these issues, we address this next.

\subsection{Choice and structure of the multiplication kernel}
\label{sec:multkern}

As explained above, the multiplication kernel is a small set
of operations repeatedly performed on each node, to build
up the rows of the product matrix $C_{i \mu , j \nu}$ for
which the node is responsible. More precisely, it is the
set of operations on which detailed coding effort will
be concentrated in order to maximise the efficiency. The choice
of kernel is therefore crucial.

Of the choices that might be made, we can envisage two
extremes, which resemble those mentioned when we
discussed the interleaving of communications and
calculations (Sec.~\ref{sec:assumptions}). At one 
extreme, we could choose the kernel to
consist of the multiplication of the $n \times n$ matrices associated
with each triplet of atoms $(i,k,j)$. (Recall that $n_1$, $n_2$,
$n_3$ are the numbers of indices on the atoms $i$, $j$, $k$; to
simplify the discussion, we refer to them without distinction as $n$.)
If we made this choice,
the kernel would sit inside a triple loop over $k$, $i$ and $j$. We refer
to this as the `fine' choice. At the other extreme is the `coarse' choice,
where the kernel consists of the entire triple loop over $k$,
$i$ and $j$, and everything inside it.
Neither extreme is likely to be satisfactory. We note that the fine choice
will ensure good reuse of data in registers, as discussed in 
Ref.~\cite{Canning96}.
In linear-scaling calculations, the dimension $n$ will be rather
small; in treating $s$-$p$ semiconductors, for example, it will
often have the value 4, so that the data for a given triplet
will generally fit into registers and primary cache. However,
each data item is used only $n$ times, and the use of
secondary cache will be poor. On the other hand, the coarse extreme
is too undifferentiated. Since the full set of $A$, $B$ and $C$ data for all
$i$, $j$, $k$ will not fit into secondary cache in any practical
situation, there is no point in making this choice of kernel. Clearly,
the operations should be broken into small groups, but these
should not be as small as $(i,k,j)$ triplets.

An obvious intermediate choice of kernel is to take it as the set
of operations associated with a given $k$. For a given $k$, the label $i$
goes over all $A$-neighbours of $k$ in the primary set, and $j$
goes over all $B$-neighbours of $k$ in the system. This will be
excellent for cache reuse of $B_{k \xi , j \nu}$, since
$j$ will run over $B$-neighbours of $k$ in the sequence in which they are
stored. The same thing will happen for $A$, provided the
$A_{i \mu , k \xi}$ elements are stored by blocks associated with
a given $k$ -- effectively this means that we store the
`local transpose' of $A$. If we denote the number of $B$-neighbours
of $k$ by $N_k^B$ and the number of primary $A$-neighbours of $k$
by $M_k^A$, then potentially every element of $A$ in cache can be
used $n N_k^B$ times, and every element of $B$ in cache can 
be used $n M_k^A$ times.

However, this still disregards cache reuse of $C$. In order
to help this, we should take the kernel to be the set of operations
associated with a small group of atoms $k$, and this group
must clearly be spatially compact. We identify this group as the
partition. A great advantage of this choice of kernel is that it fits
naturally with our proposed interleaving scheme for communications
(Sec.~\ref{sec:assumptions}). The 
overall scheme is now that each node loops over
its $A$-halo partitions. For each partition, $B$-data is communicated
for the entire partition. The multiplication kernel then
performs all the multiply-and-adds and the accumulation of $C$ for
$A$-halo atoms $k$ in the current partition.

We note in passing how the partition scheme helps efficient
data transfer from memory. The $j$ atoms that we loop over are
grouped into the spatially compact partitions used to organise the
storage of $B$. Since the $C$-matrix is also stored with the $j$-index
grouped into these partitions, and since superscalar processors
generally transfer data from main memory into cache in contiguous
sets (cache-lines)
this should considerably enhance the chances of finding
$C$-matrix elements in cache when they are needed. The underlying
thought here is that the partition scheme helps to improve
`data locality', as discussed by Goedecker~\cite{Goedecker00}.

We implied just now that, when $B$-data is fetched from the
remote node, we communicate the $B_{k \xi , j \nu}$ for {\em all}
$k$ in the current $A$-halo partition. Since these $k$ are not
necessarily all in the $A$-halo, this means that in general not
all $B_{k \xi , j \nu}$ communicated are actually used, so that
there is a waste of communications. We do this, since we believe
that it is better to tolerate the waste than to limit
communications strictly to the $B_{k \xi , j \nu}$ needed. There
are two reasons for this: first, effort would need to be put into 
generation of the indexing needed; second, the length of the
packets communicated would be reduced, so that latency would
slow the calculations.

In order to implement these ideas, we 
must now consider the two kinds of indexing
needed: first, the indexing needed to identify the atom pairs
associated with the $B$-data; second, the indexing needed to accumulate
the $C$-matrix. In order to develop an overview of the indexing
question, we first discuss schemes for labelling atoms.

\subsection{The labelling of atoms}
\label{sec:label}

\subsubsection{General ideas about labelling}
\label{sec:genlabel}

With the partition scheme outlined above, matrix multiplication
in a parallel linear-scaling code will need at least
the following five ways of labelling atoms:

\begin{enumerate}
\item
{\bf Global labelling}. As the atoms move around, and responsibilities
for atoms are passed from one node to another,
every atom in the entire system needs its own
unique label. We call this scheme `global labelling'. The natural way
is to label the atoms from 1 to $N_{\rm tot}$, where
$N_{\rm tot}$ is the total number of atoms.

\item
{\bf Partition labelling}. Since atoms are organised into partitions,
we can refer to them by saying `the $n$th atom in the $p$th partition'.
To ensure that all nodes use the same scheme, we adopt a standard
order for enumerating the partitions of the entire system, and a
standard order for the atoms in each partition -- the latter can be
in increasing order of global label, for example. We refer to this
scheme as `partition labelling'.

\item
{\bf Local labelling}. Each node needs to know only about a subset of
atoms: the atoms in its primary set, plus the atoms on other
nodes that are in range of the matrices it has to deal with.
Strictly speaking, it needs to know only about atoms in the
halo associated with the largest cut-off radius. But in order
to construct the haloes, it will need to find out about a somewhat
larger set, which we call the `grand covering set' (GCS); the number of atoms
in this set is denoted by $N_{\rm GCS}$. We label the atoms in the
GCS sequentially from 1 to $N_{\rm GCS}$, and call
this scheme `local labelling'.

\item
{\bf Halo labelling}. When looping over atoms in a particular
halo -- for example the $A$-halo when we do matrix multiplication --
it will be convenient to label the halo atoms sequentially. We refer
to this as `halo labelling'.

\item
{\bf Neighbour labelling}. In order to store matrix elements
economically, we shall want to store only those elements
$X_{i \mu , j \nu}$ etc. for which the distance between
atoms $i$ and $j$ is less than the cut-off $R_X$, since other elements
vanish by definition. For each primary atom $i$, we shall run sequentially
through $X$-neighbours $j$. This scheme for referring to neighbours
$j$ is called `neighbour labelling'. Of course, the label given
to an atom $j$ in this scheme depends on the primary atom $i$.
\end{enumerate}

Note that in these general schemes there is much freedom in the
order of the labelling. For example, in partition labelling, we can
list the partitions in different orders. Similarly, in local, halo
and neighbour labelling, we can list the atoms in different orders.
We shall want to design the storage patterns so as to optimise
the efficiency of the multiply-and-add operations, and so as
to facilitate transcription between the different labellings. To see
how this works in practice, we now study how to pass information
between nodes and how to determine the storage locations
in the $C$-array when accumulating the results of multiply-and-adds.

\subsubsection{Passing information between nodes}
\label{sec:passing}

When the matrix elements $B_{k \xi , j \nu}$ are passed
between nodes, information about the identity of the $B$-neighbours $j$
of each atom $k$ must also be passed, so that the receiving node can
determine where to accumulate the elements of the product
$C_{i \mu , j \nu}$. We have to ask what kind
of labelling should be used to send the identity of $j$.

Global labelling would clearly express the identity of $j$ unambiguously, 
but there is an objection to using this. The receiving node will need
to transcribe to the labelling it uses to refer to the storage
locations of $C$, which is essentially $C$-neighbour labelling. While
each node can transcribe from its own local, halo or neighbour
labelling to global labelling using an amount of storage that is
independent of the size of the system, transcription from global labelling
to the other kinds demands a storage proportional to the size of
the sytem, and this violates the requirement of linear-scaling
behaviour. This may not always be a problem in practice,
but it is certainly objectionable in principle.

Partition labelling offers a simple way to overcome this. For each
$B$-neighbour $j$, we specify its partition and its
sequence number in the partition, with the partition identified
by giving its offset in the three Cartesian directions relative
to the partition containing $k$. The receiving node then uses
the partition offset and sequence number to transcribe to its own
local label for each $j$. The separate indexing needed by the receiving
node to determine the storage location of $C$-elements is described
next.

\subsubsection{Determining storage locations of $C$}
\label{sec:Clocations}

To achieve the best speed, one would use a pre-calculated index array
to determine the storage location of each matrix element $C_{i \mu , j
\nu}$. As we run sequentially through the $B$-neighbours $j$ of each
atom $k$ for a given primary atom $i$, we would look up the storage
location where this contribution has to be accumulated onto the
$C$-array.  This is what was done in an earlier version of the 
{\sc Conquest} matrix-multiplication scheme~\cite{Goringe97}. The major
objection to this is that the atom $j$ in $B_{k \xi , j \nu}$ is most
easily specified in neighbour labelling, in which case the
pre-calculated index array will depend on a triplet of labels $(i, k,
j)$, so that the storage requirements of the index become exorbitant.

Our answer to this is to abandon the use of the pre-calculated index,
and instead to determine the storage locations of $C$-elements at run
time. As explained in Sec.~\ref{sec:passing}, it is straightforward to
calculate the local label of $j$ for each element $B_{k \xi , j \nu}$,
provided we pass the offset of the partition containing $j$ relative
to the partition containing $k$ in three Cartesian directions.  Our
procedure is then to transcribe from this local label to the $C$-halo
label of $j$ at run-time in the multiplication kernel, and then to
look up the address of each $C_{i \mu , j \nu}$ in a pre-calculated
index array which takes as input the primary label $i$ and the
$C$-halo label $j$. Since this index array depends on only two labels,
its storage requirements are moderate.

\subsection{Detailed coding of the multiplication kernel}
\label{sec:coding}

The principles outlined above are embodied in the F90 coding of the
multiplication kernel displayed in Fig.~\ref{fig:kernel}. This
calculates and accumulates the contributions to the $C$-matrix from
$A$-halo atoms $k$ in a given $A$-halo partition $K$, the latter being
specified in the code by the variable {\tt kpart}. Note that the outer
loop in the kernel goes only over $A$-halo atoms $k$ in $K$ (the
number of these being {\tt ahalo\%nh\_part(kpart)}), and not over {\em
all} atoms in $K$. The code inside the outer loop consists of three
sections: the three lines of the first section set up useful variables
associated with atom $k$; the purpose of the second section,
consisting of a single loop over atoms $j$, is to construct the table
{\tt jbnab2ch(j)} which transcribes from $B$-neighbour labelling to
$C$-halo labelling of the $B$-neighbour atom $j$; the third section is
a double loop over primary $A$-neighbours $i$ and $B$-neighbours $j$
of the atom $k$, within which the multiplication and accumulation are
performed. We now comment on the three sections.

In the first section, {\tt k\_in\_halo} is the $A$-halo label of
atom $k$; the array {\tt ahalo\%j\_beg(kpart)} gives us the
$A$-halo label of the first $A$-halo atom in the partition $K$.
The variable {\tt k\_in\_part} is the sequence number of atom $k$ in $K$;
this means the number in the sequence when {\em all} atoms in $K$ are
counted, and not just $A$-halo atoms. Finally, {\tt nbkbeg} is the address
where $B$-neighbours of atom $k$ start in the array {\tt b()}
holding rows of the $B$-matrix for the current partition $K$.

In the second section of the kernel, we loop over all $B$-neighbours
of the current atom $k$. The variables {\tt jpart}
and {\tt jseq} in this loop are the partition number and the
sequence number within its partition of neighbour $j$. (The
variable {\tt k\_off} is connected with periodic boundary conditions,
and will be explained in Sec.~\ref{sec:pbc}). 
The two quantities {\tt jpart}
and {\tt jseq} together consistute the `local' label of atom $j$
in the partition labelling scheme. The array {\tt jbnab2ch(j)}
transcribes from the $B$-neighbour label $j$ to the $C$-halo
sequence number, and plays a key role in what follows.

The outer loop in the third section goes over the
{\tt at\%n\_hnab(k\_in\_halo)} primary-set $A$-neighbours
$i$ of $k$, {\tt nabeg} is the address in the array {\tt a()}
where the corresponding element of $A$ is stored, {\tt i\_in\_prim}
is the sequence number of $i$ in the primary set, and {\tt icad}
is a starting address needed later. The inner loop goes over all the
{\tt nbnab(k\_in\_part)} $B$-neighbours of $k$, and {\tt nbbeg}
is the address in the array {\tt b()} holding elements of $B$.
We then use the table {\tt jbnab2ch(j)} constructed in the second
section to transcribe from $j$ to the $C$-halo label. The
pre-calculated table {\tt chalo\%i\_h2d(icad+j\_in\_halo)}
is then used to look up the address in {\tt c()} where the product
will be stored. The {\tt if} statements that test the values
of {\tt j\_in\_halo} and {\tt ncbeg} are redundant in the maximal
case $R_C \ge R_A + R_B$, since they can never be satisfied,
but they are needed for the general case, as explained later.
The triple loop over {\tt n3}, {\tt n1} and {\tt n2}
that performs the multiplication of $n \times n$ matrices
for each atom triplet $(i,k,j)$ is self-explanatory.

\subsection{Periodic boundary conditions}
\label{sec:pbc}

With periodic boundary conditions (pbc), the system we study is an
infinite Bravais lattice with a (generally large) basis. The
system is invariant under translation by an Bravais lattice vector.
For book-keeping purposes, we regard one of the primitive unit
cells of the Bravais lattice as the `fundamental' cell. The
elements of every matrix $X_{i \mu , j \nu}$ are stored for all
$i$ in the fundamental cell. Now the cut-off radius $R_X$ of any
matrix is not to be constrained in any way by the dimensions
of the repeating cell, so that the atoms $j$ for which 
$X_{i \mu , j \nu}$ is non-zero can be in the fundamental
cell or in any image of this cell. For given $i$, there can be
different atoms $j$, $j^\prime$ having non-zero values of
$X_{i \alpha , j \beta}$ which are periodic images of each other.

The use of pbc does not disturb the code structure already
established, provided one adopts the right viewpoint. Every node
knows about atoms in its primary set, and also about atoms in
its grand covering set (GCS), which contains all the haloes that it needs.
All atoms $i$ in its primary set can be regarded as belonging
to the fundamental cell, so that none is a periodic image of any
other. But atoms in its GCS may be periodic images of each
other. For most purposes, all the atoms in the GCS and
in all the haloes are regarded as completely distinct, irrespective
of whether they are periodic images or not.

To ensure efficient and correct operation of the code with pbc, two
steps must be taken. First, as we loop over $A$-halo partitions $K$,
we note that for two such partitions that are periodically 
equivalent, the sets of $B_{k \xi , j \nu}$ data for the
two partitions are identical, and we clearly do not
wish to communicate the same data more than once. We avoid this
by insisting that the order in which $A$-halo partitions
are passed through is such that any partition and its periodic
images in the halo are grouped together contiguously. Then
for each $A$-halo partition, we check whether it is periodically
equivalent to the previous one; if it is, then nothing is
communicated from the remote node, and we use the $B$-data most
recently communicated. The loop over $A$-halo partitions is outside
the multiplication kernel, so this does not affect the code shown
in Fig.~\ref{fig:kernel}.

The second step needed is to take account of the periodic offsets when
calculating the storage locations where elements of the $C$-matrix are
accumulated in the multiplication kernel. The $B$-matrix data supplied
to the kernel include a label identifying the partition to which each
$B$-neighbour atom $j$ belongs: this is given in the array {\tt
ibpart()} -- see Fig.~\ref{fig:kernel}. But this label must be
modified to account for the periodic offset of partition $K$.
Provided we use a suitable labelling of partitions in each node's GCS,
the modification consists of adding a single offset parameter {\tt
k\_off} to the partition label, this parameter being passed as an
argument to the multiplication kernel.  This addition to the first
statement in the $j$-loop (see Fig.~\ref{fig:kernel}) is the only
place in the kernel that pbc appear explicitly.

\subsection{Practical construction of indexing arrays}
\label{sec:indexing}

The multiplication kernel needs certain indexing arrays and
transcription tables, and we outline briefly the principles
used in constructing these.

In making all the labelling, a key role is played by the grand
covering set (GCS) mentioned above. In general,
a `covering set' is a set of partitions that contains all the
neighbours of an atom, or the primary-set halo, associated with a
given cut-off radius; it is used in constructing lists of neighbours
and halo atoms. The GCS is the largest such covering set, so that it can
be used for making all the neighbour and halo lists. Each node
has its own GCS. It is convenient to choose the GCS to be orthorhombic
in shape, because this simplifies the labelling. For any given
cut-off radius $R_X$, the node constructs its neighbour lists
by passing sequentially through all atoms in its primary set
and all atoms in its GCS, calculating the interatomic distances,
comparing with $R_X$, and making a record of the neighbours. This record
is then used to make the corresponding halo list.

The ordering of the lists is important, so we need to consider the
order for enumerating GCS partitions. In fact, we need to consider
two different orderings. In the first, we label GCS partitions
using the obvious Cartesian system. The GCS is orthorhombic and
the numbers of partitions along its three edges are called
$L_1$, $L_2$, $L_3$, so that any partition can be labelled by the
triplet $(l_1, l_2, l_3)$, with $1 \le l_s \le L_s$ $(s = 1,2,3)$.
Then we define the label $\lambda \equiv ( l_1 - 1 ) L_2 L_3 +
( l_2 - 1 ) L_3 + l_3$, which we call the `Cartesian composite' (CC)
label.

CC ordering is not appropriate for enumerating $A$-halo partitions
when fetching $B$-data from remote nodes, because for this we should
group partitions together by their home node and (if relevant)
according to periodic images (see Sec.~\ref{sec:pbc}). To make a
suitable ordering for this purpose, we adopt a standard order for
enumerating nodes, and a standard order for primary partitions on each
node.  Then a node enumerates the partitions of its GCS by ordering
them with respect to: (i) periodic images; (ii) partitions on each
node; (iii) node. It passes most rapidly through periodic images, so
that all images of a given partition are listed contiguously; next
most rapidly through partitions on each node, so that all partitions
on each node are contiguous; and least rapidly through nodes. It is
convenient to start with the local node itself, then to go to the next
node occurring in the GCS in increasing order of the standard node
enumeration, wrapping round in this enumeration as necessary. We call
this ordering of GCS partitions node-order periodic-grouped (NOPG).
This is the GCS ordering used for forming the $A$-halo, so that when
we pass through $A$-halo partitions we are implicitly performing a
triple loop over halo-nodes, halo-partitions on each node, and
periodic images of each halo-partition -- the latter being needed to
avoid unnecessary communication of $B$-data, as explained in
Sec.~\ref{sec:pbc}.

\subsection{The minimal case}
\label{sec:minimal}

So far, we have restricted outselves to the `maximal' case
$R_C \ge R_A + R_B$.
But, of course, we need to handle general cut-off radii.
As a first step towards doing this, we examine now the
case $R_C \le \mid R_A - R_B \mid$, which we call the `minimal'
case. We shall show that a trivial modification of the code already
presented for the maximal case gives an efficient code
for the minimal case. To see this, we must first outline
an important general principle.

For the superscalar machines that are our main interest, the speed is
limited mainly by transfers between memory and registers {\em via} the
cache: the operations performed on the data once it is in registers
have little effect on the speed\cite{Goedecker00}. So let us focus on
storage and transfer patterns, and ignore the register operations. In
the maximal case, one set of data (the $A_{i \mu , k \xi}$ matrix
elements) are in local memory and transfer occurs in a regular and
predictable way; data that behave like this will be called $X$. A
second set of data (the $B_{k \xi , j \nu}$ elements) is fetched from
remote nodes, but again is transferred in a regular way; this kind of
data will be called $Y$.  Finally, a third set of data (the $C_{i \mu
, j \nu}$ elements) is in local memory, but its transfer occurs in an
irregular way and addressing information has to be calculated for it
at run time. Data like this will be called $Z$.  The appropriate data
storage and transfer patterns for the minimal case are obtained by
identifying $C$ as $X$, $\tilde{B}$ as $Y$ and $A$ as $Z$, where
$\tilde{B}$ denotes the transpose of $B$ -- whereas in the maximal
case $A$ was $X$, $B$ was $Y$ and $C$ was $Z$. In the minimal case the
multiply-and-adds needed are $X = Z \cdot Y$ -- whereas in the maximal
case we did $Z = X \cdot Y$.  As far as the multiplication kernel is
concerned, it is transferring certain patterns of data from memory and
doing certain things with it in registers. Provided the data passed to
the kernel and the operations performed with it in registers are
changed appropriately, the maximal and minimal cases are exactly
equivalent.

It follows from this that the multiplication kernel for the minimal
case is coded exactly as in Fig.~\ref{fig:kernel}, except that the
statement within the loops over {\tt n2}, {\tt n1} and {\tt n3} must
be replaced by the following statement:
\begin{verbatim}
                    a(n3,n1,nabeg)=a(n3,n1,nabeg)+&
                      c(n1,n2,ncbeg)*b(n3,n2,nbbeg)
\end{verbatim}
The structure of all the indexing arrays remains exactly as before.
Note that in the minimal case, as in the maximal case, no
tests of cut-off distances are needed, and the {\tt if} statements
in the kernel are redundant.

\subsection{General cut-off radii}
\label{sec:cutoff}

To treat general cut-off radii, we consider first the case $R_C < R_A
+ R_B$, where $R_C$ is only a little smaller than the sum of $R_A$ and
$R_B$. This case is efficiently handled by using the code for the
maximal case (see Sec.~\ref{sec:coding}), with the $j$-loop going over
all $B$-neighbours of $k$, but with an {\tt if}-statement inserted so
that the multiply-and-adds are not done if $j$ is not a $C$-neighbour
of $i$. Needless to say, there must not be any question of calculating
interatomic distances in the multiplication kernel itself.  However,
using the arrays which transcribe from local labelling to $C$-halo
labelling, and which give the storage address in the $C$-array for
each ($i$,$j$) pair, we can determine easily whether $j$ is a
$C$-neighbour of $i$. The convention we use is that the appropriate
array elements have the value zero if the atom $j$ is not in the
$C$-halo or is not a $C$-neighbour; the test is then applied
by the two {\tt if} statements in 
Fig.~\ref{fig:kernel}. We call this way of treating the case $R_C <
R_A + R_B$ `weak reduction', the adjective `weak' referring to the
fact that it will work well only if $R_C$ does not fall too much below
$R_A + R_B$. This loss of efficiency caused by the wasted passes
through the {\tt do}-loop over $j$ is estimated in the Appendix.

If $R_C$ is very small, it will be more efficient to perform the
multiplication by `weak extension', i.e. by using the kernel for the
minimal case, but with two {\tt if}-statements inserted {\it exactly}
as for the maximal kernel (using the symmetry between the two cases
identified above in Sec.~\ref{sec:minimal}).  The effect of wasted
passes through the $j$-loop is estimated in the Appendix.

Once the {\tt if}-statements have been put into the maximal or minimal
code, any case can be treated with either kernel. We will obtain correct
results even if we treat the minimal case using the maximal kernel,
or {\em vice versa}. Nevertheless, it is clearly more efficient
to treat weak reduction using the maximal kernel and weak extension
using the minimal kernel. But how, in general, can we tell whether
it is more efficient to use the maximal or minimal kernel? This 
question is also addressed in the Appendix, where we show that the maximal
kernel should be used when $R_C > R_A$ and the minimal kernel
when $R_C < R_A$.

\subsection{Load balancing}
\label{sec:load}

However efficient the on-node operations and the communications may
be, the overall performance will be poor unless the work load is
balanced between processors, so that no processors are idle while
others are working. The partition scheme gives us a natural way to
achieve load balancing: our aim must be to group the partitions into
bundles, with each bundle containing the primary set given to a node,
in such a way that all nodes take roughly the same
computation-plus-communication time, with this time being as small as
possible. We plan to report in more detail elsewhere on load balancing
in the {\sc Conquest} code, and here we give only a brief summary of
our methods.

Our approach is to derive an approximate formula for the `cost'
of the calculation, and to use simulated annealing to search for
a way of organising partitions into bundles which comes close
to minimising the cost. This search is done by a separate code
before the matrix multiplication code is run. Let $t_i^{\rm comp}$
and $t_i^{\rm comm}$ be the times taken by node $i$ to perform
on-node computation and inter-node communication, respectively,
and let $t_i \equiv t_i^{\rm comp} + t_i^{\rm comm}$
be the sum of the two. (We assume that computation and communication
cannot be overlapped, and that the former has to wait for the latter.)
Then we have to minimise the `cost function' $T$
given by $T = \max_i ( t_i )$; this quantity $T$ is the
largest of all the $t_i$ values.

We can associate a computation time $\tau_p$ with each partition $p$.
This time is proportional to the number of $i , k , j$ triplets for which
atom $i$ is in partition $p$. The proportionality constant can be
estimated from the compute rate of the processors and the cpu efficiency,
whose determination for any processor is described in Sec.~\ref{sec:tests}.
The value of $t_i^{\rm comp}$ is then the sum of $\tau_p$ over
the partitions in the bundle assigned to processor $i$. The value of
$t_i^{\rm comm}$ is found by dividing the number of
$B_{k \xi , j \nu}$ elements that node $i$ fetches by the
practical inter-node transfer rate, with an allowance for latency.

We follow the usual methods of simulated annealing~\cite{NR}. 
Let $\Gamma$
label a given grouping of partitions into bundles -- this consists
of a list of the partitions in every bundle -- and let $T_\Gamma$
be the value of $T$ for this grouping. Then we sample the groupings
$\Gamma$ in such a way that the probability of finding $\Gamma$
is proportional to a Boltzmann function
$\exp ( - T_\Gamma / \theta )$, where $\theta$ is a fictitious
temperature. The value of $\theta$ is systematically reduced towards
zero during the simulation, so that the only surviving $\Gamma$'s
are those whose $T_\Gamma$'s are close to the minimum possible.
To sample the space of groupings $\Gamma$, we use an algorithm
to step from one $\Gamma$ to the next. As usual in simulated
annealing, the new grouping is accepted without question
if $T_\Gamma$ decreases; it is accepted with probability
$\exp ( - \Delta T_\Gamma / \theta )$ if $T_\Gamma$ increases,
where $\Delta T_\Gamma$ is the increase. 

The choice of algorithm for stepping from one $\Gamma$ to the
next is not trivial. A significant problem arises from the
rather curious definition of the cost function as the {\em maximum}
of $t_i$ values. As we search through groupings $\Gamma$, 
the time $T_\Gamma$ changes only if we make reassignments of
partitions affecting the particular bundle having the maximum
$t_i$. For large numbers of processors, this can make equilibration
of the `thermal' ensemble very slow. The measures taken
to overcome this will be described elsewhere.

\section{Coding considerations}
\label{sec:codcon}

One of the key aims when coding \textsc{Conquest} was to ensure portability
as well as performance.  Accordingly, the code was written in F90 with
all communications written using MPI1 calls (though, as explained later, 
consideration was also given to rewriting critical communications with 
high speed, system specific calls, such as {\tt shmem} on the Cray T3E).
F90 improves the readability and structure of the code through use of 
modules and derived types, while retaining enough similarity to 
\textsc{Fortran77} to be widely understood by the scientific community.

The communications are key to the efficiency of the code. There are two 
types of object to be communicated: the matrix elements themselves and the
accompanying indexing.  Efficient communication of the matrix elements 
presents no problems, since these form contiguous arrays, and are passed 
in relatively large portions (all elements for all the
atoms in a partition $K$).  The communication of the indexing, however, 
raises other problems.

The indexing for a matrix is all enclosed in a single derived type, 
{\tt type(matrix)}, of which only part needs to be communicated.  The 
derived type is defined so that it contains all the data for all the atoms
in a partition $I$ (and therefore an array of these types is defined to 
hold the data for a node's primary set).  One possibility for the 
communication of the indexing would be to use the MPI1 facilities for 
registering a new type and to pass individual partitions in a single call
for that new type.  However, the implementation of communication of defined
types within MPI1 leaves room for loss of efficiency, and another way was 
found.  Essentially, those indexing arrays which need to be communicated
are defined within the type as pointers, and point to different parts of 
an integer array.  Then the communication of all relevant indices for 
a partition can be made via a single MPI call, passing a single, large 
integer array.  Pointers are then put into place at the receiving node 
to access the indices.  Both the matrix element arrays and this large 
integer array are declared as global variables so that on the Cray 
they will be symmetric, and {\tt shmem} calls can be used on them.

While the code has been written within MPI1, the communication pattern
is naturally a one-sided one (since individual nodes require data
for partitions at disparate times).  We have developed a
quasi-one-sided scheme within MPI1, as well as using truly one-sided
schemes where available (e.g. under MPI2, which is not yet
sufficiently widely supported to be truly portable, or under {\tt
shmem} on the Cray T3E or SGI Origin2000).  At the start of the matrix 
multiplication, each node issues a series of non-buffered, non-blocking 
sends (assuming, quite reasonably, that no implementation of MPI1 would
buffer the {\tt MPI\_issend} call).  As partition data is required, nodes
issue {\tt MPI\_recv} or {\tt MPI\_irecv} calls (as appropriate) to obtain
the data.  This scheme works extremely well, and shows good performance 
and scaling.  It is easy to overlap communication and calculation by 
issuing {\tt MPI\_irecv} calls ahead of time.

\section{Practical tests}
\label{sec:tests}

We want to test the scaling properties and the efficiency of the code.
As emphasised in our earlier tests on the {\sc Conquest}
code~\cite{Goringe97}, there are different types of scaling. For given
cut-offs $R_A$, $R_B$, $R_C$, the cpu and memory needed to do the
multiplication should be proportional to the number of atoms: this is
called `intrinsic scaling'. For a given calculation on a given number
of atoms, one hopes to find cpu and memory inversely proportional to
the number of processors: `strong parallel scaling'. Finally, we can
examine the cpu and memory as the numbers of atoms and processors are
scaled up with the number of atoms per processor held fixed: `weak
parallel scaling'. There is a relation between the three scaling
types, so that we can assess the scaling behaviour completely by
studying weak and strong parallel scaling, and we shall present
results on this. Good scaling does not necessarily mean good
efficiency. For example, if the on-node operations were very
inefficient, communications could be rendered negligible, so that
parallel scaling would appear very good, even though the underlying
performance was poor. We shall therefore pay special attention to cpu
efficiency.

Our main tests are on completely random arrangements of atoms
with periodic boundary conditions; the position
of each atom in the cell is created by using a random-number generator
to draw its $x$, $y$ and $z$ coordinates from a uniform
probability distribution. We choose to use random positions
rather than, for example, perfect-crystal positions, because the
randomness should provide more of a challenge to load balancing, 
especially for small numbers of atoms per node. In order to make
contact with the real world, we take a mean density of atoms
equal to that of diamond-structure silicon, and we
use values of $R_A$, $R_B$ and $R_C$ typical of those needed in real
first-principles calculations with the the {\sc Conquest} code.
The values of the matrix elements are irrelevant for these tests,
since we are only concerned with the time taken to perform
the operations; we have actually taken their values to be a simple
radial function of interatomic separation. The number of indices
on each atom is taken to be $n = 4$.

\begin{figure}
\begin{center}
\leavevmode
\epsfxsize=100mm    
\epsfbox{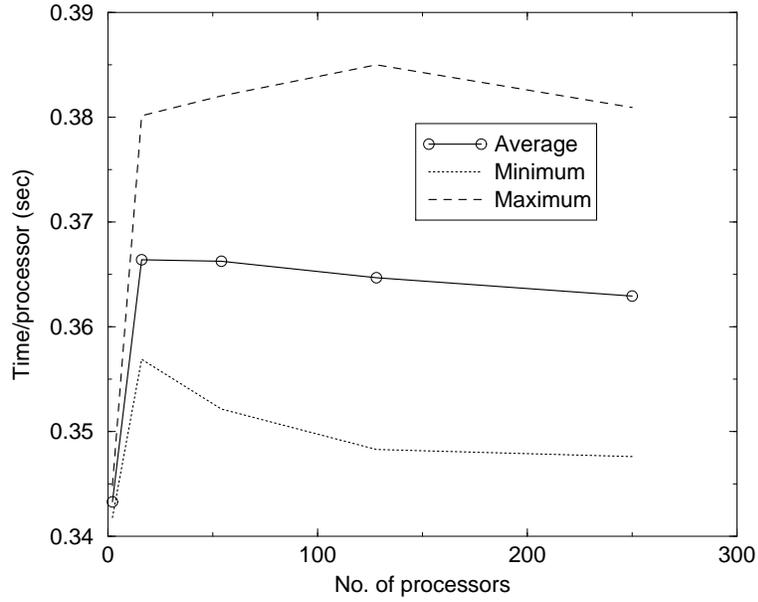}
\end{center}
\caption{Time recorded on each 
processor for the maximal
case of sparse-matrix multiplication. Numbers of processors
and atoms are varied with average number of atoms per processor 
equal to 80 in all cases.}
\label{fig:Weak}
\end{figure}

We start by presenting results obtained on the Cray T3E-1200E
using {\tt shmem} communications. We examine the maximal
case ($R_C = R_A + R_B$) with $R_A = 8.46$~\AA, $R_B = 4.23$~\AA.
This corresponds roughly to performing the multiplication
$L \cdot S$ in {\sc Conquest} ($L$ is the auxiliary density
matrix, $S$ the overlap matrix). Fig.~\ref{fig:Weak} shows results from our
tests of weak parallel scaling with 80 atoms per processor
and processor numbers going from 2 to 250 (numbers of
atoms from 160 to 20,000). The simulation cell is always a cube
in these tests, and the partitions are also taken to be cubic.
The partitions always have the same size, and the average number of
atoms per partition is 20. To obtain the timings, we put a 
timer in the code running on each node, and we report the maximum
and minimum times, together with the average time. (The times
include all communication, but they
deliberately exclude the construction of the indexing
arrays needed as input to the multiplication kernel; this is
justified, since in practice these arrays are reconstructed
only when the atoms move, whereas many matrix multiplications
are done for each set of atomic positions.) 
The following points
are noteworthy. First, the spread between minimum and maximum times
is small, so that the load balancing is good. In practice,
the time taken is governed by the time of the slowest node. Since
the maximum time is only $\sim$~6.4~\% greater than the average,
little would be gained by working harder on load balancing, at
least for random atomic arrangements. Second, the times are essentially
constant, the time on 250 nodes being only $\sim 4$~\% greater
than that on 16 nodes. This means that the size of system
that could be treated is limited only by the number of nodes
available: there is essentially no limitation from the
parallel behaviour of the code itself.

\begin{figure}
\begin{center}
\leavevmode
\epsfxsize=100mm    
\epsfbox{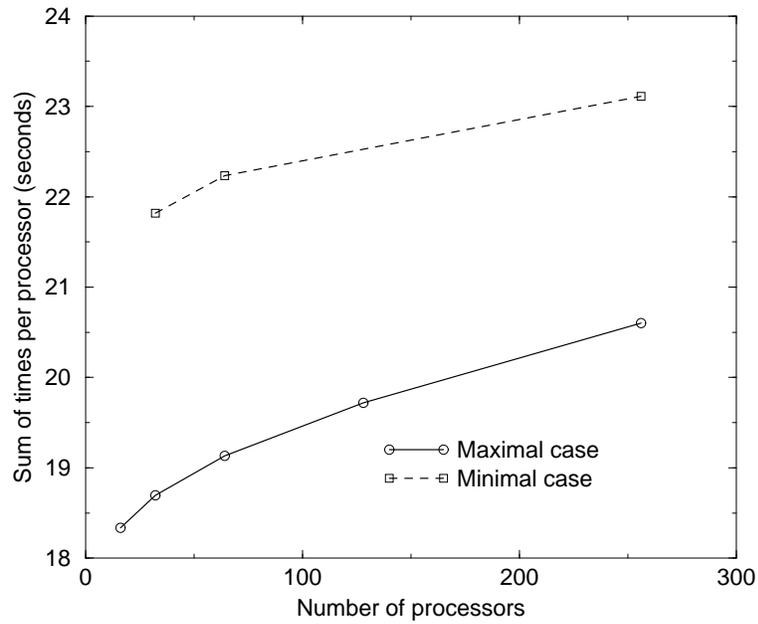}
\end{center}
\caption{Times t recorded on each node for maximal (solid line, circles)
and minimal (dashed line, squares) cases of sparse matrix multiplication 
when number of processors $N_{\rm proc}$ is varied with total number of
atoms in the system equal to 4096 in all cases.  Quantity plotted is
the sum of the times on all processors (t.$N_{\rm proc}$), which would be
exactly constant for perfect strong parallel scaling.}
\label{fig:Strong}
\end{figure}

Tests of strong parallel scaling performed for the same $R_A$ and
$R_B$ values (maximal case) with a total of 4096 atoms in the whole
cell are shown in Fig.~\ref{fig:Strong}, where we show the total time
summed over the processors as function of processor
number (which would be completely flat if the scaling were 
ideal).  A useful way to assess the scaling is by comparing the times
with Amdahl's law, which states that the time taken is $t = t_0 ( ( 1
- x ) + x / N_{\rm proc} )$, where $N_{\rm proc}$ is the number of
processors, with $x$ and $( 1 - x )$ the fractions of the calculation
performed in parallel and serially. This formula fits the maximum
times very well, and we find the value $x = 0.9986$, so that the
serial fraction is 0.0014. This means that strong scaling is
extremely good up to $\sim 200$ processors, or $\sim 20$ atoms
per processor. Since it would be wasteful of memory to run {\sc
Conquest} with fewer atoms per processor than this, the implication is
that strong scaling will be very well satisfied in practical
calculations.

We now turn to the minimal case ($R_C \le \mid R_A - R_B \mid$).  We
argued in Sec.~\ref{sec:minimal} that the speed will be limited mainly
by transfers to and from memory, rather than by operations in
registers. This implies that, provided we use the minimal, rather than
the maximal, multiplication kernel, the timings for the multiplication
with $R_A = 12.69$~\AA, $R_B = 4.23$~\AA, $R_C = 8.46$~\AA\ should be
almost the same as for $R_A = 8.46$~\AA, $R_B = 4.23$~\AA, $R_C =
12.69$~\AA. We have tested this explicitly by performing
multiplication with $R_A = 12.69$~\AA, $R_B = 4.23$~\AA, $R_C =
8.46$~\AA\ on 4096 atoms with different numbers of processors. The
results are also reported in Fig.~\ref{fig:Strong}. We see that the
timings are very close to our results for the maximal case, though
higher by $\sim 16$~\%. This means that all our conclusions about weak
and strong scaling will be the same for the minimal case.

We now turn to cpu efficiency. Here, we are solely concerned
with the execution speed of the multiplication kernel itself,
so that communication time is excluded. As explained in the
Introduction, efficiency is defined in terms of the minimum
number of operations required to perform the multiplications
of eqn~(\ref{eqn:mult}). For a given node, this is the
number of $(i,k,j)$ triplets for $i$ in that node's primary
set, multiplied by $n_1 \times n_3 \times n_2$, multiplied
by a factor of 2 to allow for the fact that we do a multiply
and add for each $( i \mu , k \xi , j \nu )$. This  minimum
number of operations divided by the time spent in the multiplication
kernel gives the `useful' Mflop rate, which is then divided
by the peak Mflop rate of the processor to give the
cpu efficiency.

We have done efficiency tests on both perfect-crystal and random
atomic positions. As an illustration, we report results on
perfect-crystal silicon (lattice parameter~= 5.46~\AA) with cut-offs
$R_A = 6$~\AA, $R_B = 10$~\AA, $R_C = 16$~\AA\ with 1728 atoms in the
total system. The tests were done with 64 cubic partitions and 16
processors on the Cray T3E-1200E, the partitions being distributed
unevenly over the processors. The `useful' Mflop rate was found to be
almost identical on all processors, the average being 101.4~Mflops,
and minimum and maximum being 101.1 and 103.6~Mflops. The theoretical
peak speed of the processor is 1.2~Gflops, so that the efficiency is
8.4~\%. This result is reasonably satisfactory in two ways. First, it
shows that all the operations associated with transcription between
labellings, the determination of storage addresses, etc cannot be
taking a large amount of time (recall that these are not included in
the number of `useful' operations); second, they show that fairly good
use is being made of the cache.

\begin{figure}
\begin{center}
\leavevmode
\epsfxsize=100mm    
\epsfbox{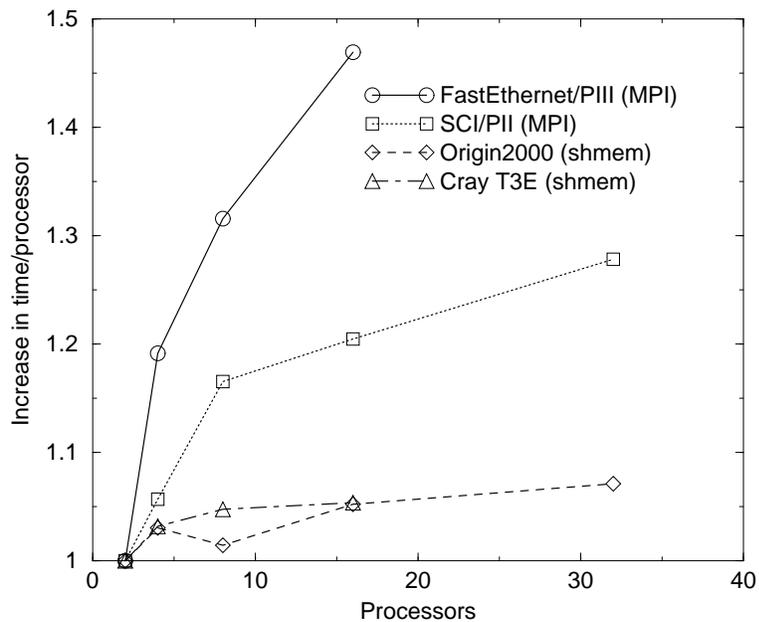}
\end{center}
\caption{Average time per processor as a multiple of time on two processors
for constant number of atoms per processor (equal to 32).  Maximal case for 
crystalline silicon with cut-off radii $R_A$ = 12\AA, $R_B$ = 8\AA and 
$R_C$ = 20\AA.}
\label{fig:systems}
\end{figure}

We have also performed tests on a number of other machines: a Beowulf
linked by fast ethernet; a Beowulf linked by SCI interconnect; and a SGI
Origin2000.  We have tested weak scaling on 
all these systems to investigate the portability of the code, and 
the importance of efficient communications.  The {\em increase} in
time per node as the system size is increased is shown in 
Fig.~\ref{fig:systems}.  We remark that, given perfect communication, 
the line ought to be flat at 1.0, and that to all intents and purposes
this is true for the Origin2000 and T3E.  The SCI beowulf system performs
well and should be adequate for {\sc Conquest} calculations.  The fast 
ethernet beowulf, however, shows far too large an increase.

Finally, we assess briefly the implications of our new methods
for practical calculations with the full {\sc Conquest} code.
The present matrix multiplication code has been incorporated
into {\sc Conquest}, and we have run tests in which the energy of
the self-consistent ground state is calculated for
silicon systems. In searching for the ground state, the
basic step is the `self-consistency iteration', in which an
input charge density is used to calculate the Kohn-Sham
Hamiltonian matrix, and the techniques of Ref.~\cite{Bowler99} are
used to determine the $L$-matrix that yields the non-self-consistent
ground state for the current Hamiltonian, and hence the output
charge density.  We search for self-consistency using the 
GR-Pulay technique\cite{Bowler00b}.

\begin{figure}
\begin{center}
\leavevmode
\epsfxsize=100mm    
\epsfbox{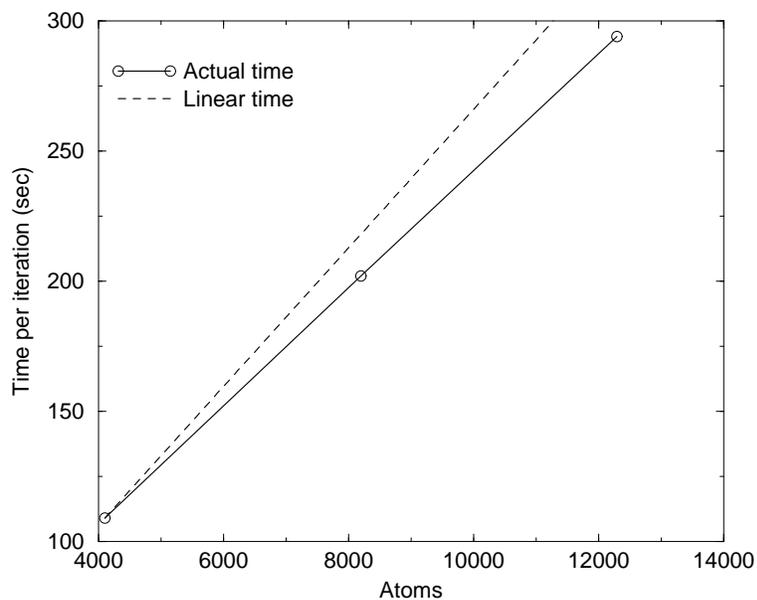}
\end{center}
\caption{Time for one self-consistency iteration (see text) of
the {\sc Conquest} code as a function of number of atoms of
perfect-crystal silicon. Calculations were run on 512 processors
of a Cray T3E-1200E. Points connected by solid line show elapsed
time in seconds; dashed line
shows time scaled linearly from the actual time for 4096 atoms.}
\label{fig:Conquest}
\end{figure}

We report in Fig.~\ref{fig:Conquest} timings for a single
self-consistency iteration obtained for crystalline silicon containing
numbers of atoms going from 4096 to 12288 when the calculations are
run on 512 processors of the T3E-1200E.  For these tests, we used a
norm-conserving non-local pseudopotential\cite{Kleinman82}.  The
support-region radius $R_{\rm reg}$ was equal to 2.715~\AA\ and the
$L$-matrix cut-off, $R_L$ equal to 6.9~\AA\ (for definitions of these
cut-offs, see Ref.~\cite{Hernandez96}); these values are typical of
what would be used in practice. The Figure shows that the scaling with
system size is so good that the time is actually a slightly sub-linear
function of atom number. We believe that this is because, as the
primary sets become larger, the number of nodes with which
communication is required actually decreases, making the time required
decrease.  To put these results in context, we note that, starting
from scratch, typically 10 self-consistency iterations are needed to
reach the self-consistent ground state with acceptable accuracy for
fixed support functions, and that a further factor of 10 should be
allowed for variation of the support functions. This means that, with
the methods presented here, the attainment of the full ground state
with an accuracy comparable to that expected in a conventional
plane-wave calculation can be accomplished for a 12,000-atom system in
a few hours on a 512-processor T3E.

\section{Discussion and conclusions}
\label{sec:disc}

We have shown that our code for parallel multiplication of
locally sparse matrices achieves two important things:
first, it displays almost perfect linear scaling with respect
to numbers of atoms and processors; second, it achieves good
cpu efficiency.

We remark that we have every right to expect the linear scaling
to be perfect when the numbers of processors and atoms
are increased together, with a constant number of atoms per
processor (weak scaling): for this type of scaling, the amounts
of computation and communication should, indeed, remain
constant. What is less trivial is the type of scaling in which
the number of processors is increased with the total number of
atoms held fixed (strong scaling). In contrast to
some previous schemes, the present code shows excellent strong
scaling, at least on machines having a fast communications
network. Good cpu efficiency is also non-trivial.

A key idea in the present scheme is the management of the atoms
in small compact groups, referred to as `partitions'. Although
this certainly makes the code more complex, it has the enormous
advantage of giving an internal structure to the calculation,
which can be used in a number of ways: it gives a flexible means
of constructing the primary sets that are assigned to processors;
it enables us to manage the communications by breaking them into
packets of convenient size; it does the same for the on-node
computations, so that we work with a convenient `multiplication kernel';
it provides a labelling scheme which allows nodes to
identify to each other the atoms for which data is communicated;
finally, it aids efficient cache use by improving data locality.
It is important to note that the present scheme of partitions
and bundles for managing atoms
strongly resembles the scheme of blocks and domains for managing
integration-grid points in {\sc Conquest} described in our
earlier work~\cite{Goringe97}. A further key concept in the present
work is that of transcription between different atom-labelling
schemes, which is crucial in reducing the computational
overheads in the multiplication kernel.

At the start of our analysis (Sec.~\ref{sec:assumptions}), we made
certain assumptions about the storage of matrix elements
and the way the matrix multiplication should be done, and we
now ask whether anything could be gained by changing these
assumptions. Our answer is that little is likely to be gained.
With the scheme we have presented, the scaling properties are already
close to ideal, and there is little room for improvement. It is
true that our cpu efficiency is only $\sim 10$~\%, and there
is clearly scope for improvement here. However, one should
remember that efficiencies as high as 50~\% on superscalar
processors are achievable only in the multiplication
of large non-sparse matrices, and that serious sparsity is
bound to incur a serious loss of efficiency. Our suspicion
is that an improvement of the cpu efficiency by as much as a factor
of two is unlikely, and that any such improvements are more likely
to come from work on the multiplication kernel than by changing
the basic assumptions. Time will tell.

\section*{Acknowledgments}
\label{sec:ack}

We would like to thank Dr.~S.~Goedecker for allowing us to read an 
early copy of Ref.~\cite{Goedecker00}.
The work of DRB was supported by EPSRC grant GR/M01753 and by an EPSRC
Postdoctoral Fellowship in Theoretical Physics.
The position of MJG is partially
supported by GEC and CLRC Daresbury Laboratory. Calculations on the
Cray T3E at CSAR and on the Origin 2000 at the UCL HiPerSPACE Centre
were supported by grants GR/M01753 and JR98UCGI. We are indebted to
Dr. S.~Pickles and colleagues at CSAR for their help in optimising the
code. Useful discussions with Dr.~I.~Bush in the early stages of the
work are acknowledged.  Assistance from Dr.~B.~Searle, Mr.~R.~Harker 
and Mr.~A.~Keller with tests on beowulf systems is also acknowledged.

\section*{Appendix: Reduction and extension, maximal
and minimal kernels}

As explained in the text, different multiplication kernels
should be used for large and small values of the cut-off
radius $R_C$ of the product matrix $C$: for large and small $R_C$, we
should use the maximal and minimal kernels respectively.
In this Appendix, we discuss how to decide the cross-over
point from the maximal to the minimal case
as the radii $R_A$, $R_B$ and $R_C$ are varied.

If we have particular values of $R_A$, $R_B$ and $R_C$,
then for each atom $i$ in a node's primary set, there is a certain
number of pairs $(k,j)$ for which multiply-and-adds have to
be done. Our algorithm for deciding the cross-over point will
be based on estimates for this number of pairs, which
we denote by $\nu_{\rm p}$. If we treat the problem as weak reduction,
then we ignore the constraint supplied by $R_C$, and instead of
the true number of pairs $\nu_{\rm p}$ we actually visit
a larger number of pairs, which we call $\nu_{\rm p}^{\rm max}$.
The merit of doing it like this can be gauged by the ratio
$\eta^{\rm max} \equiv \nu_{\rm p} / \nu_{\rm p}^{\rm max}$.
If this ratio becomes
too small, then it is wasteful to treat it as weak reduction. Similarly,
if we treat as weak extension, we visit a number of pairs
$\nu_{\rm p}^{\rm min}$ which will generally be greater than
$\nu_{\rm p}$, so we can gauge the merit of this method by the
ratio $\eta^{\rm min} \equiv \nu_{\rm p} / \nu_{\rm p}^{\rm min}$.
We should go for the method that gives the largest $\eta$ value,
so the policy will be that for given values of $R_A$,
$R_B$ and $R_C$ we treat it as weak reduction if $\nu_{\rm p}^{\rm max} <
\nu_{\rm p}^{\rm min}$ and as weak extension if $\nu_{\rm p}^{\rm min} <
\nu_{\rm p}^{\rm max}$.

The values of $\nu_{\rm p}$, $\nu_{\rm p}^{\rm max}$ and
$\nu_{\rm p}^{\rm min}$ in practice will depend on the details
of the atomic positions. We need a simple approximate way of
estimating the numbers of pairs. To do this, we assume
that the density of atoms is uniform, with
$\rho_0$ atoms per unit volume. Then our estimate for the number of
pairs is:
\begin{equation}
\nu_{\rm p} = \rho_0^2
\int_{r_1 < R_A , \, r_2 < R_C , \, \mid {\bf r}_2 - {\bf r}_1 \mid < R_B}
d {\bf r}_1 d {\bf r}_2
\end{equation}
It is simple geometry to evaluate the integral. Let $\Omega ( R_A ,
R_B ; r )$ be the overlap volume of two spheres
of radius $R_A$, $R_B$ when the distance between their
centres is $r$. Then:
\begin{equation}
\nu_{\rm p} ( R_A , R_B , R_C ) = 4 \pi \rho_0^2
\int_0^{R_C} dr \, r^2 \Omega ( R_A , R_B ; r ) \; .
\label{eqn:volume}
\end{equation}

It will be useful to note some symmetry properties of
$\nu_{\rm p} ( R_A , R_B , R_C )$. Since $\Omega ( R_A , R_B ; r )$
is invariant under interchange of $R_A$ and $R_B$, the
same is true of $\nu_{\rm p}$. In fact, it is easy to
show that $\nu_{\rm p}$ is completely invariant with respect
to all permutations of $R_A$, $R_B$ and $R_C$:
\begin{eqnarray}
\nu_{\rm p} ( R_A , R_B , R_C ) & = & \nu_{\rm p} ( R_C , R_A , R_B ) =
\nu_{\rm p} ( R_B , R_C , R_A ) \nonumber \\
& = & \nu_{\rm p} ( R_B , R_A , R_C ) = \nu_{\rm p} ( R_A , R_C , R_B ) =
\nu_{\rm p} ( R_C , R_B , R_A ) \; .
\end{eqnarray}

These symmetries very much simplify the calculation of $\nu_{\rm p}$
for general values of $R_A$, $R_B$ and $R_C$. Let $R_1$, $R_2$ and
$R_3$ denote the smallest, the middle and the largest of $R_A$,
$R_B$ and $R_C$.
Then for arbitrary values of $R_A$, $R_B$ and $R_C$, we can
write:
\begin{equation}
\nu_{\rm p} ( R_A , R_B , R_C ) = \nu_{\rm p} ( R_1 , R_2 , R_3 ) \; .
\end{equation}
There are two separate cases that must be considered, which we
refer to as `triangle' and `non-triangle'. In the first
case, we have $R_3 \le R_1 + R_2$, so that the three lengths
can form the sides of a triangle. In the second, we have
$R_3 > R_1 + R_2$, and they cannot form the sides of a triangle.
The non-triangle case is trivial, because the constraint
supplied by $R_3$ is of no effect, and we just have:
\begin{equation}
\nu_{\rm p} ( R_1 , R_2 , R_3 ) = \rho_0^2
\int_{r_1 < R_1 , \, r_2 < R_2}
d {\bf r}_1 d {\bf r}_2 \,    = \frac{16}{9} \pi^2 \rho_0^2
R_1^3 R_2^3 \; .
\label{eq:pair_nont}
\end{equation}
In the `triangle' case, we return to eqn.~(\ref{eqn:volume})
and note that the overlap volume of two spheres is:
\begin{eqnarray}
\Omega ( R_1 , R_2 ; r ) & = & \frac{4}{3} \pi R_1^3 \; \; \; \;
{\rm if} \; \; \; r < R_2 - R_1 \nonumber \\
& = & - \frac{\pi}{4 r} ( R_2^2 - R_1^2 )^2 +
\frac{2}{3} \pi ( R_1^3 - R_2^3 ) -
\frac{1}{2} \pi r ( R_1^2 + R_2^2 ) +
\frac{\pi}{12} r^3 \; \; \; \; \; {\rm if} \; \; \; r > R_2 - R_1 \; .
\end{eqnarray}
The integral over $r$ is then straightforward, and we get:
\begin{eqnarray}
\nu_{\rm p} & = & \pi^2 \rho_0^2 \left( 
\frac{16}{9} R_1^3 ( R_2 - R_1 )^3
- \frac{1}{2} ( R_2^2 - R_1^2 )^2 \left[ R_3^2 - ( R_2 - R_1 )^2 \right]
+ \frac{8}{9} ( R_1^3 + R_2^3 ) \left[ R_3^3 - ( R_2 - R_1 )^3 \right]
\right.
\nonumber \\
& & \left. \mbox{} - \frac{1}{2} ( R_1^4 + R_2^4 )
\left[ R_3^4 - ( R_2 - R_1 )^4 \right]
+ \frac{1}{18} \left[ R_3^6 - ( R_2 - R_1 )^6 \right] \right) \; .
\label{eq:pair_t}
\end{eqnarray}

It is now simple to find the cross-over point and the $\eta$ value
at this point. The strategy for weak reduction is based
on ignoring the constraint supplied by $R_C$. Then $\nu_{\rm p}^{\rm max}$
is just $\nu_{\rm p}$ for the non-triangle case $R_C > R_A + R_B$,
so that from eqn.~(\ref{eq:pair_nont}) we have:
\begin{equation}
\nu_{\rm p}^{\rm max} = \frac{16}{9} \pi^2 \rho_0^2 R_A^3 R_B^3 \; .
\end{equation}
In weak extension, we ignore the constraint supplied by $R_A$. Then
$\nu_{\rm p}^{\rm min}$ is $\nu_{\rm p}$ for the
non-triangle case $R_A > R_B + R_C$, so we have:
\begin{equation}
\nu_{\rm p}^{\rm min} = \frac{16}{9} \pi^2 \rho_0^2 R_B^3 R_C^3 \; .
\end{equation}
The cross-over point is the point at which $\nu_{\rm p}^{\rm max} =
\nu_{\rm p}^{\rm min}$, which requires that $R_A = R_C$.
In terms of the dimensionless ratio $\alpha = R_C / ( R_A + R_B )$,
we can express the cross-over point as:
\begin{equation}
\alpha_{\rm X} = 1 / ( 1 + \xi ) \; ,
\end{equation}
where $\xi = R_B / R_A$.

The efficiency $\eta_{\rm X} = \nu_{\rm p}^{\rm max} / \nu_{\rm p} =
\nu_{\rm p}^{\rm min} / \nu_{\rm p}$ at the cross-over point
can now be found from eqn.~(\ref{eq:pair_t}). The derivation of the
resulting formula:
\begin{equation}
\eta_{\rm X} = 1 - \frac{9}{16} \xi + \frac{1}{32} \xi^3
\end{equation}
is straightforward.

In practice, we should always try to ensure that $R_A \ge R_B$,
so that $0 < \xi < 1$. It is readily shown that $\eta_{\rm X}$
increases montonically from $15/32$ to 1 as $\xi$ decreases
from 1 to 0. This means that the efficiency cannot drop
below $15/32$, or, in round numbers, 50~\%.

\begin{figure}
\begin{center}
\begin{verbatim}
do k=1,ahalo%nh_part(kpart)  ! Loop over atoms k in current A-halo partn
  ! --- useful variables associated with atom k ---------------------
  k_in_halo=ahalo%j_beg(kpart)+k-1
  k_in_part=ahalo%j_seq(k_in_halo)
  nbkbeg=ibaddr(k_in_part)
  ! --- transcription of j from B-neighbour to C-halo labelling -----
  do j=1,nbnab(k_in_part)
    jpart=ibpart(nbkbeg+j-1)+k_off
    jseq=ibseq(nbkbeg+j-1)
    jbnab2ch(j)=chalo%i_halo(chalo%i_hbeg(jpart)+jseq-1)
  enddo
  ! --- perform multiply and adds -----------------------------------
  do i=1,at%n_hnab(k_in_halo)  ! Loop over primary-set A-neighbours of k
    nabeg=at%i_beg(k_in_halo)+i-1
    i_in_prim=at%i_prim(at%i_beg(k_in_halo)+i-1)
    icad=(i_in_prim-1)*chalo%ni_in_halo
    do j=1,nbnab(k_in_part) ! Loop over B-neighbours of atom k
      nbbeg=nbkbeg+j-1
      j_in_halo=jbnab2ch(j)
      if(j_in_halo/=0) then
        ncbeg=chalo%i_h2d(icad+j_in_halo)
        if(ncbeg/=0) then  ! multiplication of ndim x ndim blocks 
          do n2=1,ndim3
            do n1=1,ndim1
             do n3=1,ndim2
                c(n1,n2,ncbeg)=c(n1,n2,ncbeg)+ &
                  a(n3,n1,nabeg)*b(n3,n2,nbbeg)
              enddo
            enddo
          enddo
        endif
      endif ! End of if(j_in_halo.ne.0)
    enddo ! End of 1,nbnab
  enddo ! End of 1,at%n_hnab
enddo ! End of k=1,ahalo%nh_part(kpart)
return
\end{verbatim}
\end{center}
\caption{The matrix multiplication kernel.}
\label{fig:kernel}
\end{figure}

\end{document}